\documentclass[journal,letter,10pt]{IEEEtran}

\ifCLASSOPTIONcompsoc
    \usepackage[caption=false, font=normalsize, labelfont=sf, textfont=sf]{subfig}
\else
\usepackage[caption=false, font=footnotesize]{subfig}
\fi
\usepackage{graphicx, epstopdf}
\usepackage{lipsum}
\usepackage{amssymb}
\usepackage{amsmath}
\usepackage{amsthm}
\usepackage{mathrsfs}
\usepackage{cite}
\usepackage{color}
\usepackage[acronym]{glossaries}
\usepackage[T1]{fontenc}
\usepackage[latin9]{inputenc}
\usepackage{array}
\usepackage{textcomp}
\usepackage{multirow}
\usepackage{flushend}
\usepackage{xcolor}
\usepackage{slashbox}

\title{Collaborative RF and Lightwave Power Transfer for Next-Generation Wireless Networks}

\author{ Ha-Vu Tran, Georges Kaddoum, and Chadi Abou-Rjeily
\thanks{
Ha-Vu Tran and Georges Kaddoum are with LACIME Laboratory, University of Qu\'{e}bec, \'{E}TS engineering school, 1100 Notre-Dame west, H3C 1K3, Montreal, Canada. Emails: \{ha-vu.tran.1@ens.etsmtl.ca, georges.kaddoum@etsmtl.ca.\}

Ha-Vu Tran is also with Duy Tan University, Danang, Vietnam.

Chadi Abou-Rjeily is with the Department of Electrical and Computer Engineering of the Lebanese American University (e-mail: chadi.abourjeily@lau.edu.lb).

}
}

\begin{document}
    \maketitle
\begin{abstract}
Breakthroughs in information and power transfer for wireless networks are imperative in order to satisfy the requirement of wireless nodes for energy sustainability. To this end, significant research efforts in academia and industry have been devoted to the design of optimal resource allocation schemes for RF simultaneous wireless information and power transfer (SWIPT) networks. The transmit power constraints imposed by safety regulations introduce significant challenges to the improvement of the power transfer performance. Therefore, solely relying on RF resources to cope with the expectations of next-generation wireless networks, such as longer device lifetimes and higher data rates, may no longer be possible. Thus, the investigation of  technologies complementary to conventional RF SWIPT is of critical importance. In this article, we propose a novel collaborative RF and lightwave power transfer technology for future networks, where both the RF and lightwave bands can be entirely exploited. In this context, we introduce the basic transceiver architecture and four corresponding collaborative communication and power transfer protocols. Finally, key potential future research directions are highlighted.
\end{abstract}

\section{Introduction}
The rise of the information and communication technology (ICT) era has led to the massive deployment of wireless devices in diverse networks, including the fifth generation (5G) cellular, Internet-of-Things (IoT), and E-healthcare networks \cite{MahmoudKamel}. Generally, a substantial burden is on these devices, where their constant provision of wireless access and the fact that they are battery operated, and thus energy-constrained, severely affect their operational lifetime. Replacing the battery may seem like a simple task; however, given the ever-increasing number of devices and their limited accessibility in some applications (e.g., in hazardous environments), battery replacement is often operationally expensive or infeasible.

To cope with this issue, simultaneous wireless information and power transfer (SWIPT) technology has been proposed, which enables wireless devices to be continuously recharged via electromagnetic waves \cite{BrunoClerckx,Diamantoulakis2018}, and opens new possibilities for the sustainability of future networks. SWIPT based on transmitted radio frequency (RF) waves is an interesting solution; however, due to the inevitable spectrum crunch, solely relying on RF may not be possible due to the need to jointly meet the expectations of (i) 10-fold longer device battery lifetimes \cite{AGupta2015}, and (ii) the global mobile data traffic in excess of 100 exabytes \cite{Ericsson}, while respecting the transmit power constraints imposed due to various safety concerns \cite{WHO238}. 

Recently, optical wireless communication (OWC) has emerged as a complementary technology to RF for the provisioning of indoor wireless links \cite{Pathak2015}. On the platform of OWC, wireless energy recharging over lightwave, i.e., visible light (VL) and near infrared light (NIRL), has recently attracted a surge of attention from both academia and industry. The high potential of lightwave power transfer techonology has been confirmed in several previous works \cite{Fakidis,GaofengPan,Diamantoulakis2018}. However, this technology may only be used if a line-of-sight (LoS) is available in order to transfer a sufficient amount of power for energy harvesting (EH). In addition, VL is constrained by the requirement to maintain constant illumination while NIRL is restricted by eye- and skin-safety conditions \cite{safetylaser}. 

It is apparent from their relative advantages and disadvantages that RF and lightwave wireless power transfer approaches are complementary rather than competitive technologies. Since the combination of lightwave and RF can entirely exploit the efficacy of both the RF and lightwave bands for wireless power transfer (WPT), it is possible to retain RF pollution lower than the safe limit while increasing the WPT performance. Therefore, the novel combination and the joint optimization of the RF and lightwave power transfer has the potential to efficiently provide safe and reliable WPT technology for next-generation wireless networks.

This article proposes the applications of RF and lightwave power transfer to indoor wireless communication networks. Its core contents are described as follows: First, a brief overview of the current RF and lightwave-based WPT techniques is presented and the associated health and safety concerns are discussed. We then present a novel framework for collaborative RF and lightwave power transfer, consisting of a transceiver architecture with four communication and power transfer protocols. Finally, future research directions are comprehensively discussed.

\section{RF and Lightwave Power Transfer and Related Health Concerns} 
Although both RF and lightwave communications have revolutionized our world, they can have detrimental effects on the health of individuals and hence must be deployed with care. In this section, we provide a brief overview of RF and lightwave power transfer technologies, and discuss some related health concerns.

\subsection{RF Wireless Power Transfer}
RF WPT is a technique that wirelessly delivers electrical energy over RF signals to recharge the devices in next-generation cellular, wireless body area, and IoT networks. Deploying RF WPT systems introduces several new concerns for human health due to RF radiation. In RF WPT systems, the sensitivity of an EH receiver, i.e., $-10$ dBm, is much higher than that of a conventional information decoding (ID) receiver, i.e., $-60$ dBm (see \cite{BrunoClerckx} and references therein). To enable RF WPT, manufacturers and network operators need to increase the transmit power levels to achieve a reasonable EH performance at the receiver side. In addition, since RF WPT systems are often deployed in buildings where people could be stationary or moving, human bodies may unexpectedly block the transmission. This not only causes a significant EH performance loss at the intended devices but may also result in adverse human health effects due to exposure to RF fields \cite{WHO238}. These effects include localized heating or stimulation of excitable tissue.
More specifically, according to the IEEE C95.1-2005 Standard, regarding the continuous and long-term exposure of individuals to RF signals, the fundamental restriction of the specific absorption rate (SAR) for frequencies between $100$ kHz and $3$ GHz is $0.08$ W/kg, averaged over any $6$-minute-reference-period \cite[pp. 20]{IEEEstandard}. Concretely, assuming a body mass of $60$ kg, there is a SAR limit of $4.8$ W averaged over a period of $6$ minutes that can be applied on a human body. This limit guarantees that there is no risk of any adverse effects on human health.  

Although the RF WPT approach has gained much support from the industry over the past few years, the safety concern may have not received full consideration yet.
However, it is obvious that addressing the human safety concerns results in a strict limit for the wireless energy beamed to devices. Moreover, in practical scenarios, the actual harvested energy might be much lower than the beamed energy due to energy conversion losses. 
Thus, novel technologies must be considered for the realistic implementation of WPT in future networks.

\subsection{Lightwave Power Transfer: Visible and Infrared Light}
Inspired by the high potential of OWC technology, lightwave WPT and EH from artificial light are attracting increasing attention. 
There are two main research directions of lightwave power transfer: using VL and using NIRL. 

In the case of visible lightwave power transfer, which uses the $430 - 770$ THz spectrum band, the achievable EH performance in indoor visible light communication (VLC) networks is quite limited. The reason behind this is twofold: {(i)} the intensity of LED lights is lower than that of solar light and LED bulbs in buildings are not always turned on, and { (ii)} the illumination in indoor living environments should be from $200$ luminous flux (lx) to $1000$ lx to respect the eye safety standard \cite{Europeanlighting}. Meanwhile, the main advantage of VLC is that the indoor light brings {\it free} energy since there is no extra transmit power needed from the lighting system.

\begin{figure*}[t]
\centering
\fbox{\includegraphics[width=0.75\textwidth]{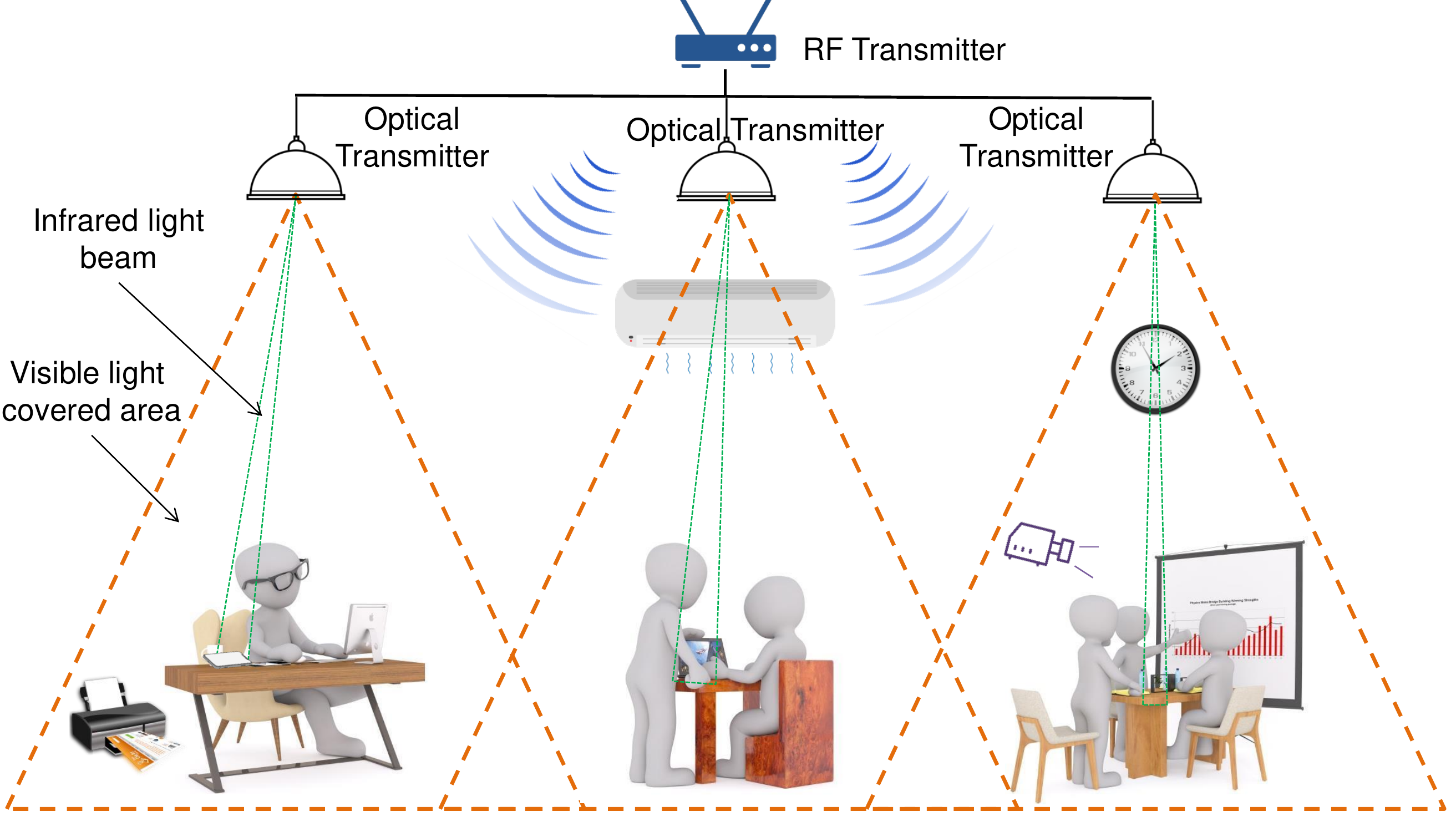}}
  \caption{
      An illustration of the proposed collaborative RF and lightwave power transfer network.
    }
\label{fig:1}
\end{figure*}

On the other hand, the near infrared spectrum (300 - 400 THz band) causes no light pollution to human eyes, and therefore is less constrained. Nonetheless, incautiously handling the light emission in this band could cause injuries to humans. Thus, network operators need to obey the safety standard of the irradiance on eyes and skin, i.e., for example, $0.005$ W/m$^2$ at the wavelength $905$ nm for long-time exposures \cite{safetylaser}. 
This limitation is general for NIRL-applied wireless networks and not particularly related to NIRL power transfer.
Meanwhile, since NIRL does not cause light pollution to human eyes; to improve the efficiency of EH, the optical transmitter can form very narrow light beams with high intensity over the safety standard but must keep the light beams away from the human body. 
Particularly, NIRL can be interrupted when the link is blocked by using the distributed resonating laser method \cite{QLiu2016}, which can diminish any possible health risks.
Indeed, the high potential of this approach has been demonstrated. 
Furthermore, since the wavelength of NIRL is in the order of hundreds of nanometers and the size of the photodetector is in the order of centimeters, the detector size is sufficiently large to fully absorb NIRL signals. Therefore, if the NIRL beam is narrow enough to focus the light within the photodetector area, we can neglect the reflections of NIRL from the surrounding surfaces of the devices directed to human eyes. In fact, unlike the VL radiation pattern that is designed to be broad for the sake of realizing efficient illumination, NIRL beams are very narrow in order to achieve efficient information transfer.

In light of the above discussion, the combination of RF and lightwave power transfer is important, even when only using RF is sufficient, to not provoke any health risks.
To this end, it can be seen that the lightwave and RF WPT are complementary technologies. In the next section, we will discuss how we can combine them to enable novel safe WPT for next-generation wireless networks.

\section{Enabling Collaborative RF and Lightwave Power Transfer }
Since RF, VL, and NIRL do not interfere with each other, they may be applied to WPT in a mutually supportive way. In this section, we present a novel architecture for collaborative RF and lightwave power transfer networks as well as four distinct communication and power transfer protocols for different scenarios.

\subsection{ Collaborative RF and Lightwave Power Transfer Architectures}
The combination of RF and lightwave power transfer can fully exploit the multi-band efficacy of RF, VL, and NIRL for WPT, while satisfying the safety induced transmit power constraints per band. Fig. \ref{fig:1} illustrates a collaborative RF and lightwave power transfer network scenario where terminal devices can accumulate the energy harvested from light and RF.

\begin{figure}[h]
\centering
\fbox{\includegraphics[width=0.38\textwidth]{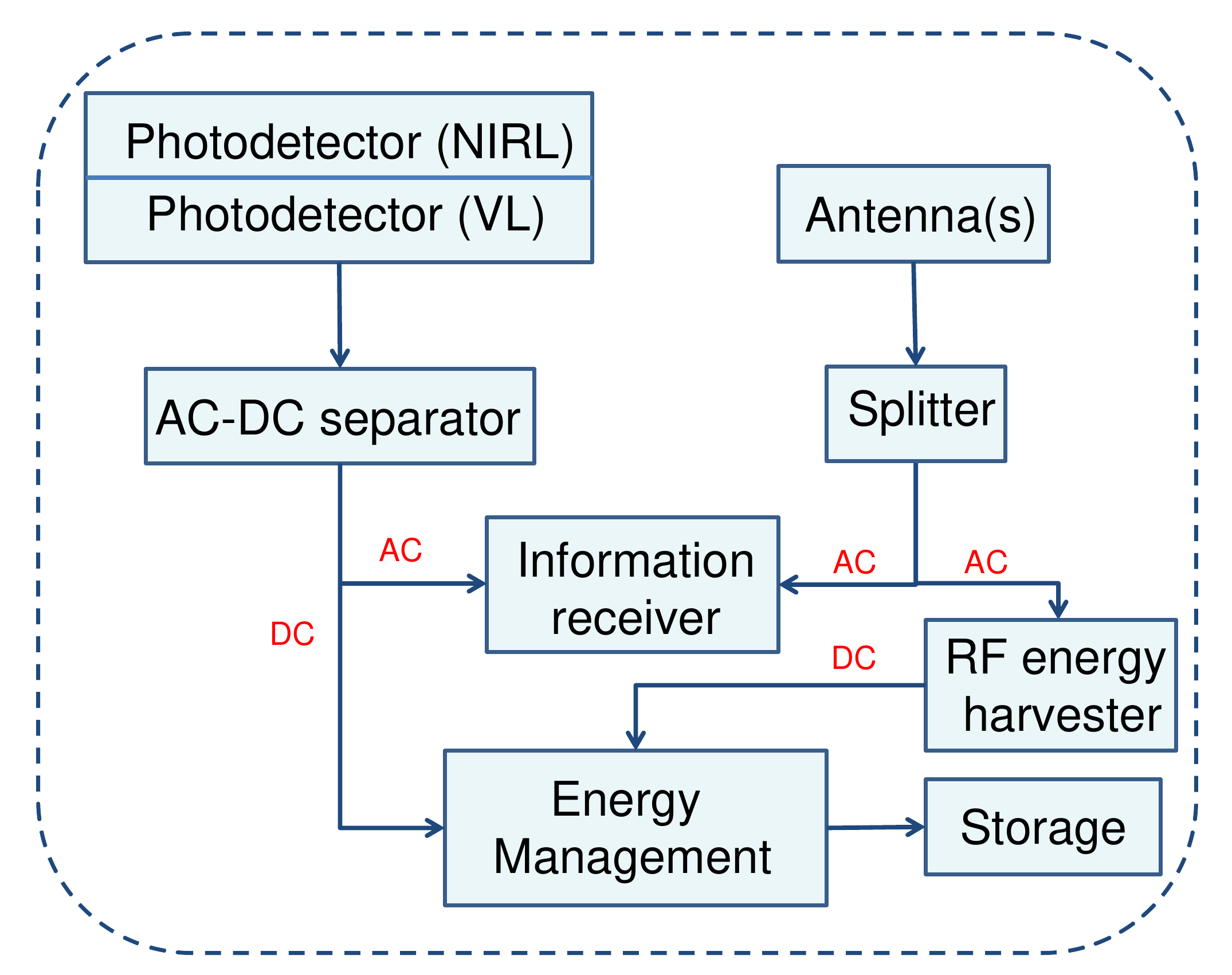}}\\
\caption{Proposed receiver architecture.}
\label{fig:transceiver}
\end{figure}

The transmitter architecture, based on a joint design of optical and RF transmitters, is shown in {Fig. \ref{fig:1}. In most cases, each optical transmitter, generally an LED bulb, only has a micro-controller. {Therefore, all computational tasks are processed at the RF transmitter, which is often equipped with a powerful digital baseband unit}.

With the existing technology, it is not possible to harvest energy from VL and NIRL using the same photodetector. Hence, we propose a receiver architecture using two photodetectors in the photovoltaic mode, as shown in Fig. \ref{fig:transceiver}. Since the photodetector capturing NIRL can be transparent \cite{Husain2018} and allows the VL to pass through, it can be installed above the VL photodetector.
In this architecture, two different photodetectors are responsible of receiving data and energy from the VL and NIRL, respectively, which are then conveyed to an alternating current (AC)-to-direct current (DC) separator in order to extract the DC component from the light signals for the EH purpose while the AC component is used for the ID purpose. Since NIRL and VL do not interfere with each other, only one AC-to-DC separator is needed. The utilization of photovoltaic mode is simple and energy efficient for both ID and EH. In fact, EH realized from converting the AC component to DC component (based on rectifiers) is complex and not energy efficiency since it involves active circuits \cite{Pan2019}.

Besides, in the architecture, the RF signals are captured by antenna(s) and then sent to a splitter where they are divided into two streams for ID and EH purposes, respectively.  The part for ID is conveyed to the information receiver, while the one for EH is directed to the EH unit where it is converted to DC form. Based on the harvested energy from both RF and light and concurrent energy consumption, the energy management module decides either to draw energy from the storage or convey excessive energy to the storage for future use.

The lightwave ID-EH performance management at the transmitter can be performed in the power domain and the time domain using power splitting and time switching methods, respectively \cite{Diamantoulakis2018}. The power splitting method is based on controlling the AC-to-DC ratio to leverage the ID and EH performances. On the other hand, the time switching method is based on dividing each time frame into two slots. In this context, the time switching method aims to maximize the ID performance during the first time-slot by setting the AC component at its maximum admissible level. In the second time-slot that is fully dedicated to EH, the target is to maximize the EH by maximizing the DC component.

Different from the lightwave ID and EH management, the handling of the RF ID and EH is performed at the receiver. In this context, RF signals, received from antennas, are split into two parts for ID and EH. According to ID and EH performance requirements, an optimal ratio to split the received RF signals is derived. 

\begin{figure*}[!]
\centering
\fbox{\includegraphics[width=0.9\textwidth]{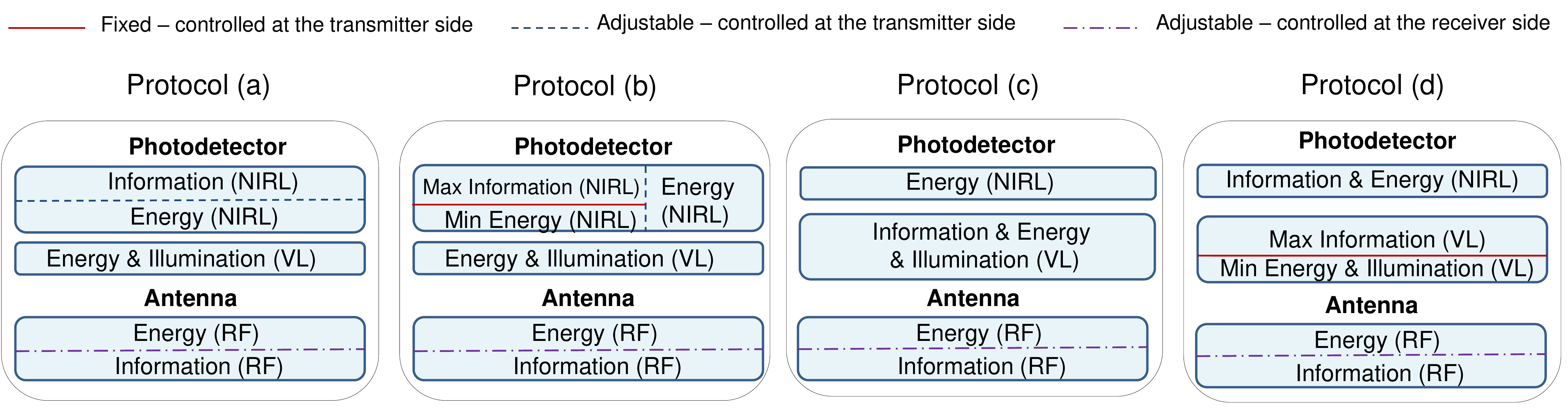}}
  \caption{
       Information and power transfer protocols.
    }
    \label{fig:protocol}
\end{figure*}

\begin{table*}
\begin{centering}
\caption{Protocol Characteristics}
\par\end{centering}
\centering{}%
\begin{tabular}{|c|c|c|c|c|}
\hline
\backslashbox{\textbf{Characteristics}}{\textbf{Protocols}} &
\textbf{ (a) } &
\textbf{ (b) } &
\textbf{ (c)} &
\textbf{ (d)} 
\tabularnewline
\hline
NIRL & Power splitting & Time switching  & Max Energy & Power Splitting/Time Switching
\tabularnewline
\hline
VL & Max Energy & Max Energy & Power Splitting/ Time Switching & Max Information \& Min Illumination
\tabularnewline
\hline
RF &  Power Splitting &   Power Splitting &  Power Splitting &  Power Splitting
\tabularnewline
\hline
Power efficiency &  No &   No &  Yes &  No
\tabularnewline
\hline
Dark environment &  Not supported &   Not supported &  Not supported &  Supported
\tabularnewline
\hline
EH performance &  Very high &   Very high &  Very high&  High
\tabularnewline
\hline
ID performance &  Very High &   Very High &  High&  Very high
\tabularnewline
\hline
\end{tabular}
\end{table*}

Up to now, there are some products which can act as external receivers, enabling the reception of information and energy from VL and NIRL for conventional RF terminal devices. Besides, there have been some efforts to integrate light energy harvesters into such devices. Therefore, it is firmly believed that receiver architectures enabling lightwave and RF information and power transfer are feasible and will be realized in the near future.

\subsection{Information and Power Transfer Protocols}
Indeed, there are many possible combinations of using VL, NIRL, and RF according to network requirements. However, to maximize the total power transfer performance while maintaining information transmission, our idea is to allocate the primary function of delivering information and power to NIRL, while the RF and the VL act as complementary wireless power sources. It can be explained by that NIRL is able to provide significantly higher data rates than RF \cite{Pathak2015}. Additionally, the NIRL band is more convenient than the VL band for transferring energy because there are no constraints on consistent lighting in covered areas.  Moreover, RF is very useful in case of non-line-of-sight (NLOS) since VL and NIRL signals are blocked by obstacles. In this context, motivated by the fact that the ID and EH performances can be managed in the time and power domains, we first introduce two preliminary protocols, namely (a) and (b), as follows.

{{\it Protocol (a):}} A power splitting-based collaborative protocol is shown in Fig. \ref{fig:protocol}, where RF delivers information and energy, NIRL conveys both information and energy, and VL maintains the illumination in the served area and also contribute to the EH performance. In this context, AC and DC-biased components of the NIRL signals determine the ID and EH performance, respectively. Thus, managing the ID and EH performances can be achieved by adjusting the ratio between the AC and DC powers.

{{\it Protocol (b):}} NIRL carries both information and power whereas VL conveys energy and illumination. The use of RF in this protocol is similar to that in {protocol (a)}. Particularly, the management of the ID and EH associated to the NIRL relies on time switching, as shown in Fig. \ref{fig:protocol}. In this context, the time frame is divided into two time-slots, for information transmission and power transfer, where the performances of information and energy are maximized, respectively. Particularly, since NIRL signals always include a DC-biased component; the light energy can also be harvested in the time-slot dedicated to information transmission.

\begin{figure*}[t]
\centering
\subfloat[Rate-Energy region of each individual approach]{\emph{\fbox{\includegraphics[width=0.45\textwidth]{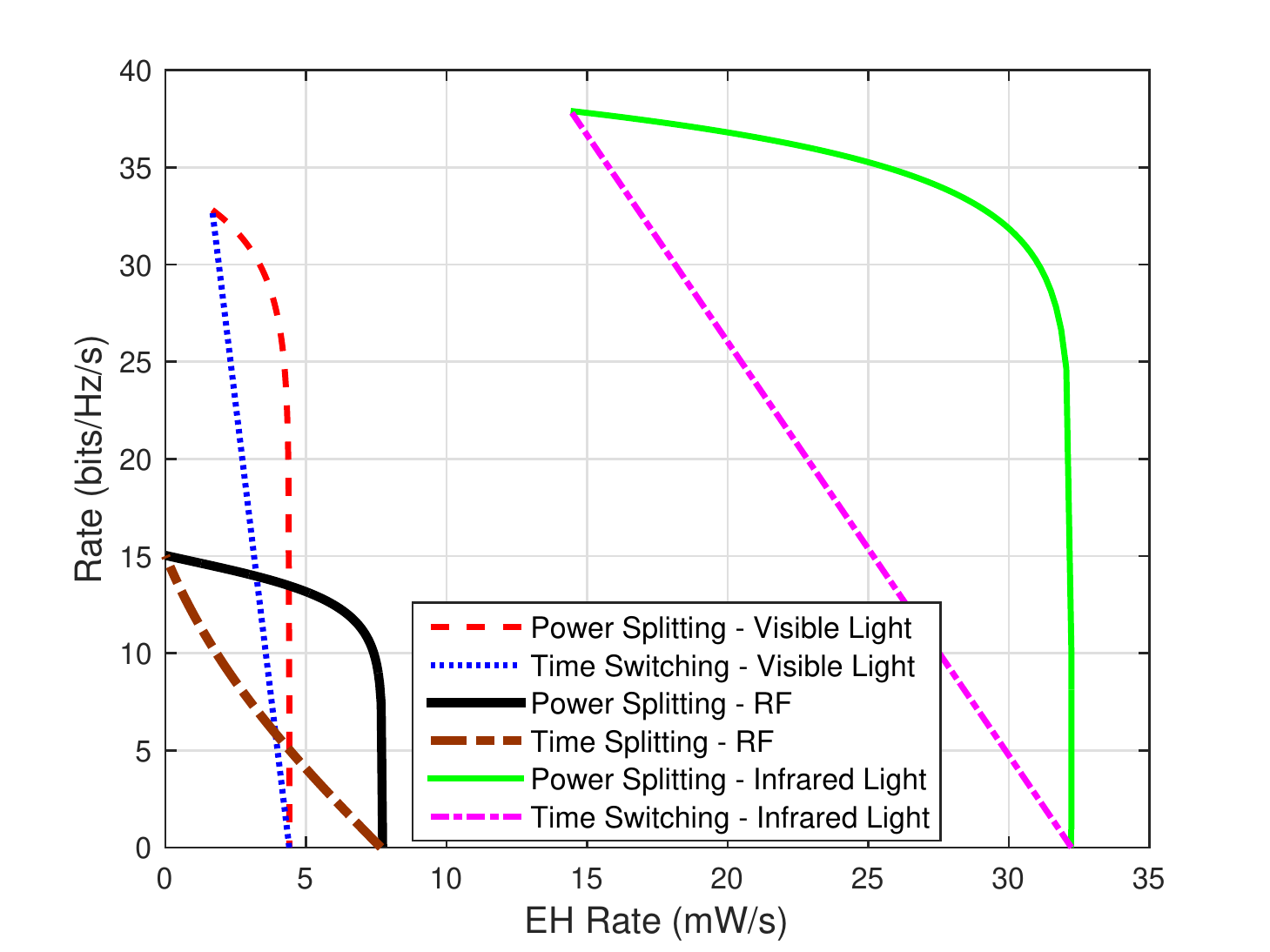}}}} \quad
\subfloat[Rate-Energy region of the proposed protocols.]{\emph{\fbox{\includegraphics[width=0.45\textwidth]{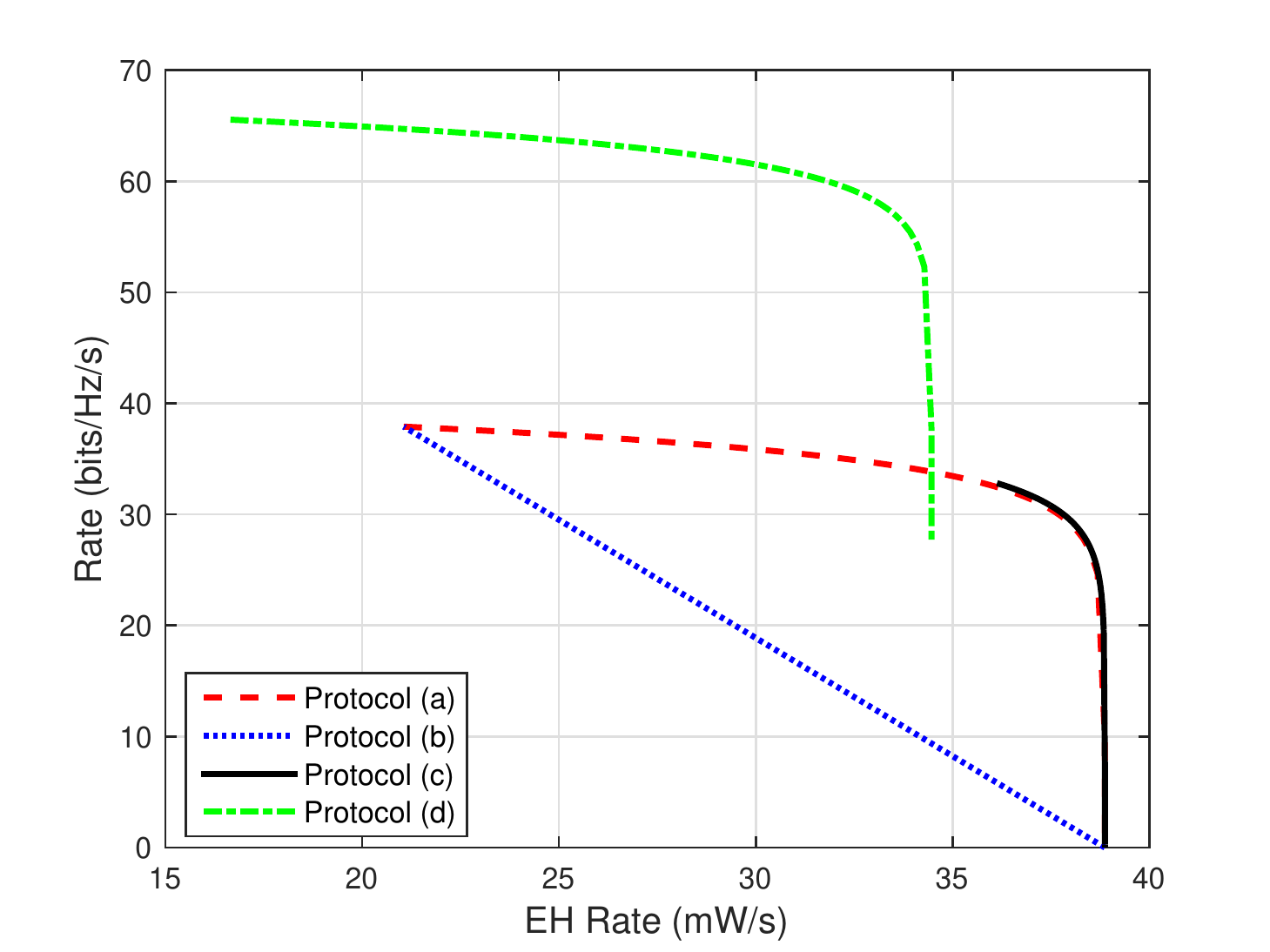}}}}
\caption{Scenarios of collaborative RF and lightwave power transfer. }
\label{fig:sim}
\end{figure*}

Furthermore, on the basis of protocols (a) and (b), we then present two more protocols (c) and (d) to address the specific cases where the served areas have strong background light and dimmed light, respectively, in the following.

{{\it Protocol (c):}} This protocol, presented in Fig. \ref{fig:protocol}, is designed for the sake of transmit power efficiency.
The main motivation behind this protocol is due to the fact that although NIRL can deliver data at very high rates as VL, it consumes additional transmit power. 
However, as mentioned, no extra transmit power is needed from the lighting system for VL to convey either information or energy.
Hence, in protocol (c), we propose releasing NIRL from the task of information transmission. More specifically, NIRL is in charge of transferring power while the role of VL and RF are similar to that given in protocols (a) and (b).

{{\it Protocol (d):}} In contrast with {protocol (c)}, this protocol addresses the scenario of dimmed light in the served areas, e.g., a bedroom during rest time. The use of VL, in this case, is set in a dimmable configuration that only focuses on maximizing the ID performance with low-intensity light. This is mainly due to the fact that although the light intensity can be enough to ensure a reasonable data rate, it might be insufficient for harvesting energy. In this protocol, NIRL and RF play a complementary role to improve the ID and EH performances whenever necessary. Also, NIRL can be handled by either the power-splitting or time switching manner.

Note that in the proposed protocols, the ratio of AC to DC components and the RF power-splitting factor can be optimized according to the various network requirements. 

Given the proposed transmitter and receiver architectures, one can see that the task of handling the lightwave ID and EH performances is conducted at the transmitter while managing the RF ID and EH ones is performed at the receiver, similar to the conventional approach \cite{BrunoClerckx}. Therefore, the processing complexity at the receiver is reasonable and comparable with the existing literature. This renders the four proposed protocols implementable in practice.

\subsection{Performance of Collaborative RF and Lightwave Power Transfer}
We simulate a network consisting of a four-antenna RF transmitter, an optical transmitter, and three single-antenna terminal devices equipped with two solar pannels. 

We aim to investigate the rate-energy regions of the RF, VL, and NIRL schemes. In this regard, the distances from the RF transmitter and the optical transmitter to all the terminal devices, e.g. an IoT device, are $4$ m and $2.05$ m, respectively. In this simulation, the RF channel is subject to the Rician fading with a Rician factor of 6 dB and a pathloss exponent of 2.6. Besides, the maximal ratio transmission (MRT) beamformer is used and a total transmit power of $20$ dBm is equally allocated among all devices.
Further, for convenience, we assume that the optical transmitter delivers the same power to each terminal device. The optical channel parameters in this article are similar to those in \cite{Diamantoulakis2018}. The served area has a normal lighting level where the LED bulb's power is $22$ W with semi-angle at half power equal to $60^\circ$. The NIRL angle-diversity bulb's power is $66$ W with a semi-angle at half power equal to $15^\circ$ for each angle element. Further, the incident angle of light radiation is $60^\circ$, and the optical band-pass filter gain of transmission is equal to 1. For convenience, regarding NIRL and VL photodetectors, the photodetector areas are $85$ cm$^2$ (i.e. the phone screen size), the photodetector responsivities are $0.4$, the fill factors are $0.75$, and the noise power is $10^{-15}$.
In principle, the rate-energy region of VL can be derived by investigating, in an exhaustive manner, all possible rate-energy trade-offs when adjusting the ratio between the AC and DC components of optical signals, and the time fractions for ID and EH. In this context, the corresponding data rate and EH performances are calculated using the rate formula for lightwave and the optical EH model \cite{Diamantoulakis2018}, respectively. Similarly, the rate-energy region of RF can be obtained based on modifying the RF signal portions used for ID and EH. In this context, the RF data rate and RF EH metrics are computed according to the conventional rate formula and the nonlinear EH model \cite{BrunoClerckx} (i.e. $P_{Sat} = 24$ mW, $a = 150$, $b=0.014$; with the same notations as in \cite{BrunoClerckx}), respectively.

As observed in Fig. \ref{fig:sim}a, VL and NIRL can offer a much higher data rate than RF. Nevertheless, RF outperforms VL when considering the EH performance. 
Further, the achievable performance of the four proposed protocols is depicted in Fig. \ref{fig:sim}b.
In general, the combined use of RF, VL, and NIRL expands the rate-energy region significantly compared to the sole use of either of the three resources. In Fig. \ref{fig:sim}b, the RF transmit power used for WPT is $16$ dB, implying less RF radiation compared with the simulation given Fig \ref{fig:sim}a. However, the achievable EH of any protocol still outperforms the EH of any individual approach in Fig. \ref{fig:sim}a. In fact, the collaborative RF and lightwave power transfer approach allows us to reduce the RF radiation while maintaining a decent EH performance.

\section{Future Research Opportunities}
Since RF, VL, and NIRL are complementary to each other, their combination can be applied to many interesting applications in next-generation networks. In the following, we discuss some future research opportunities.

\subsection{Indoor Applications}
\subsubsection{Secure Wireless Powered Communication Networks} In addition to the benefits of data rate, wireless energy transfer, and illumination; the application of the proposed collaborative approach can also offer high security at the physical layer of wireless networks. In other words, the combination of RF, VL, and NIRL can be extensively exploited to design new secure WPT scenarios to better protect information while transfering data and wireless energy to devices.
Specifically, this is motivated by the fact that light cannot pass through opaque structures, therefore the information that it carries is secured from users in other rooms or buildings. However, even though security is one of the main advantages of OWC, dealing with the circumstance where eavesdroppers exist in the covered network area, i.e., in the same room, is challenging.  For instance, we assume that eavesdroppers are able to detect optical and RF signals. If the eavesdroppers are passive, the design of angle-diversity optical transmitters with color allocation may help improve the network security, whereas RF can play a role as artificial noise to decrease the eavesdropping performance while enhancing the EH performance at the device. If the eavesdroppers are active and jam the network using either RF or NIRL (they may not use VL to jam the network since this manner will be able to reveal their position), the network may use VL to secure communications while jamming back the eavesdropper using RF and NIRL. As a consequence, the EH performance is improved.

\subsubsection{E-Healthcare}
With the development of ICT, doctors can diagnose patients remotely. Thus, data transferred from medical devices to the doctors should be accurate and timely because it might be critical to the lives of patients. The use of the lightwave for downlink can replenish energy for medical or security devices such as handheld medical instruments, sensors, and smart locks for health monitoring, and indoor localization. Indeed, this technology is a promising candidate for E-Healthcare, particularly in Telemedicine and Mobile Health (mHealth) applications where patients can be connected with doctors and nurses in hospitals, at home, or anywhere through one secure platform. Further, in these indoor applications, one of the main challenges is providing high quality services while accounting for human eye safety regulations and maintaing consistent illumination in served areas.

\subsection{Outdoor Applications}
\subsubsection{Unmanned aerial vehicles}
In disaster situations, such as an earthquake, providing continuous wireless communication is vital for rescuing victims. In such situations, unmanned aerial vehicles (UAVs) could act as a relay to convey information as well as to recharge devices over the air. A group of UAVs can maintain communication and simultaneously recharge some devices, while searching for victims in devastated buildings. In this regard, RF can be used to provide continuous transmission for alert and command signals, which are not in need of high data rate communication. Otherwise, lightwave is employed.
Regarding wireless power transfer, in the presence of LoS transmission, since narrow light beams can be created more conveniently than narrow RF beams, the lightwave power transfer is preferred. Otherwise, RF is a more suitable option. In this scenario, optimizing the position of the group of UAVs to improve energy transfer efficiency is one of the most crucial challenges. Further, the health safety of victims and rescue staffs should be guaranteed while delivering wireless power.

\subsubsection{Underwater Communication and Power Transfer}
While RF is an efficient approach for terrestrial communications, its use for underwater communications faces many difficulties due to dramatic absorption losses. Generating RF signals for underwater communications requires a considerable antenna size and a tremendous transmit power, namely in hundreds of Watts, to achieve a $100$ Mbps data rate. 
Further, the use of RF for wireless power transfer is very challenging.
On the other hand, narrow light beams can be generated at a low cost and only need several Watts to achieve Gbps data rates. Therefore, lightwave outperforms RF for underwater applications. 
However, when both the transmitter and receiver are moving, this scenario might entail a big challenge in terms of dealing with pointing errors due to bore-sight and jitter effects.
Hence, a hybrid scheme, where lightwave is used to transmit power and information at high data rates over a shorter range whereas acoustic technology is employed to send command and control signals (i.e., not requiring high data rates) over the longer range,  is a promising future direction for this field.

\section{Conclusion}
In this article, we presented a brief overview of RF and optical power transfer techniques. Furthermore, we introduced a framework for collaborative RF and lightwave power transfer. In this regard, we developed a transceiver architecture suitable for hybrid RF and optical WPT. Moreover, four communication and power transfer protocols for distinct scenarios were proposed and challenged through simulations. We showed that supplemental performance gains can be achieved by intelligently combining RF and lightwave technologies.  Finally, several future research directions have been highlighted to foster continuous advancement in this area. It is firmly believed that this will become a fruitful research topic in the very near future.

\section{Acknowledgement}
The authors would like to deeply thank Prof. Robert Schober, Prof.  George K. Karagiannidis, and Dr. Panagiotis D. Diamantoulakis for their valuable comments on this work.

\bibliographystyle{IEEEtran}
\bibliography{IEEEabrv,REF}

\begin{thebibliography}{10}
\providecommand{\url}[1]{#1}
\csname url@samestyle\endcsname
\providecommand{\newblock}{\relax}
\providecommand{\bibinfo}[2]{#2}
\providecommand{\BIBentrySTDinterwordspacing}{\spaceskip=0pt\relax}
\providecommand{\BIBentryALTinterwordstretchfactor}{4}
\providecommand{\BIBentryALTinterwordspacing}{\spaceskip=\fontdimen2\font plus
\BIBentryALTinterwordstretchfactor\fontdimen3\font minus
  \fontdimen4\font\relax}
\providecommand{\BIBforeignlanguage}[2]{{%
\expandafter\ifx\csname l@#1\endcsname\relax
\typeout{** WARNING: IEEEtran.bst: No hyphenation pattern has been}%
\typeout{** loaded for the language `#1'. Using the pattern for}%
\typeout{** the default language instead.}%
\else
\language=\csname l@#1\endcsname
\fi
#2}}
\providecommand{\BIBdecl}{\relax}
\BIBdecl

\bibitem{MahmoudKamel}
M.~Kamel, W.~Hamouda, and A.~Youssef, ``Ultra-dense networks: A survey,''
  \emph{IEEE Comm. Surveys \& Tutorials}, vol.~18, no.~4, pp. 2522 -- 2545,
  Fourthquarter 2016.

\bibitem{BrunoClerckx}
B.~Clerckx, R.~Zhang, R.~Schober, D.~W.~K. Ng, D.~I. Kim, and H.~V. Poor,
  ``Fundamentals of wireless information and power transfer: From {RF} energy
  harvester models to signal and system designs,'' \emph{{IEEE} J. Sel. Areas
  Commun.}, vol.~37, no.~1, pp. 4--33, Jan. 2019.

\bibitem{Diamantoulakis2018}
P.~D. Diamantoulakis, G.~K. Karagiannidis, and Z.~Ding, ``Simultaneous
  lightwave information and power transfer (slipt),'' \emph{IEEE Transactions
  on Green Communication and Networking}, vol.~2, no.~3, pp. 764--773, Sept.
  2018.

\bibitem{AGupta2015}
A.~Gupta and R.~K. Jha, ``A survey of {5G} network: Architecture and emerging
  technologies,'' \emph{IEEE Access}, vol.~3, pp. 1206 -- 1232, July 2015.

\bibitem{Ericsson}
\BIBentryALTinterwordspacing
``Mobile data traffic growth outlook,'' Ericsson Mobility Report, Tech. Rep.,
  Nov. 2017, accessed on Nov. 7th, 2019. [Online]. Available:
  \url{https://www.ericsson.com/en/mobility-report/reports}
\BIBentrySTDinterwordspacing

\bibitem{WHO238}
\emph{Extremely Low Frequency Fields, Environmental Health Criteria 238}, World
  Health Organization (WHO) Std. ISBN 9\,789\,241\,572\,385, 2007.

\bibitem{Pathak2015}
P.~H. Pathak, X.~Feng, P.~Hu, and P.~Mohapatra, ``Visible light communication,
  networking, and sensing: A survey, potential and challenges,'' \emph{IEEE
  Comm. Surveys \& Tutorials}, vol.~17, no.~4, pp. 2047 -- 2077, Fourth quarter
  2015.

\bibitem{Fakidis}
J.~Fakidis, S.~Videv, S.~Kucera, H.~Claussen, and H.~Haas, ``Indoor optical
  wireless power transfer to small cells at nighttime,'' \emph{IEEE/OSA J.
  Lightw. Technol.}, vol.~34, no.~13, pp. 3236--3258, Jul. 2016.

\bibitem{GaofengPan}
G.~Pan, J.~Ye, and Z.~Ding, ``Secure hybrid {VLC-RF} systems with light energy
  harvesting,'' \emph{{IEEE} Trans. Commun.}, vol.~65, no.~10, pp. 4348 --
  4359, Otc. 2017.

\bibitem{safetylaser}
\emph{Safety of laser products - Part 1: Equipment classification, requirements
  and user's guide}, IEC Std. 60\,825-1, Aug. 2001.

\bibitem{IEEEstandard}
\emph{IEEE Standard for Safety Levels With Respect to Human Exposure to Radio
  Frequency Electromagnetic Fields, 3 k{H}z to 300 {GH}z}, IEEE Std.
  C95.1-2005, Otc. 2005.

\bibitem{Europeanlighting}
\emph{Light and lighting -- Lighting of work places -- Part 1: Indoor work
  places}, the European lighting standard -- EN12464-1:2011 Std.

\bibitem{QLiu2016}
Q.~. Liu, J.~Wu, S.~Z. P.~Xia, W.~Chen, Y.~Yang, and L.~Hanzo, ``Charging
  unplugged: {WillDi} stributed laser charging for mobile wireless power
  transfer work ?'' \emph{IEEE Veh. Technol. Mag.}, vol.~2, no.~4, pp. 36--45,
  Dec. 2016.

\bibitem{Husain2018}
A.~A. Husain, W.~Z.~W. Hasan, S.~Shafie, M.~N. Hamidon, and S.~S. Pandey, ``A
  review of transparent solar photovoltaic technologies,'' \emph{Renewable and
  Sustainable Energy Reviews}, vol.~94, pp. 779--791, 2018.

\bibitem{Pan2019}
G.~{Pan}, P.~D. {Diamantoulakis}, Z.~{Ma}, Z.~{Ding}, and G.~K.
  {Karagiannidis}, ``Simultaneous lightwave information and power transfer:
  Policies, techniques, and future directions,'' \emph{IEEE Access}, vol.~7,
  pp. 28\,250--28\,257, 2019.

\end{thebibliography}

\end{document}